\documentstyle[prd,aps,floats,tabularx,epsfig,comment]{revtex}

\newcommand{\ETAL}{{\it et al.}}
\newcommand{\be}{\begin{equation}}
\newcommand{\ee}{\end{equation}}
\newcommand{\ba}{\begin{eqnarray}}
\newcommand{\ea}{\end{eqnarray}}
\newcommand{\hld}{\hspace{0.25cm}\cdots} 
\newcommand{\fns}{\footnotesize}
\newcommand{\mbs}[1]{\mbox{\small #1}}
\begin{document}

\draft
\preprint{\
\begin{tabular}{rr}
&
\end{tabular}
}
\twocolumn[\hsize\textwidth\columnwidth\hsize\csname@twocolumnfalse\endcsname
\title{The initial conditions of the universe: how much isocurvature is allowed?}
\author{M.~Bucher$^1$, J.~Dunkley$^2$, P.~G.~Ferreira$^2$, 
K.~Moodley$^{2,3}$, C.~Skordis$^2$}
\address{$^1$DAMTP, Centre for Mathematical Sciences, University of 
Cambridge, Wilberforce Road, \\ Cambridge CB3 0WA, United Kingdom \\
$^2$Astrophysics, University of Oxford, 
Denys Wilkinson Building, Keble Road, \\ Oxford OX1 3RH, United Kingdom \\
$^3$School of Mathematical Sciences,
University of KwaZulu-Natal, Durban, 4041, South Africa }
\maketitle
\vskip -0.15in
\begin{abstract} 

We investigate the constraints imposed by the current data on correlated 
mixtures of adiabatic and non-adiabatic primordial perturbations. We 
discover subtle flat directions in parameter space that tolerate 
large ($\sim 60\%$) contributions of non-adiabatic fluctuations. In 
particular, larger values of the baryon density and a spectral tilt 
are allowed. The cancellations in the degenerate directions are
explored and the role of priors elucidated.
\end{abstract}

\date{\today}
]
\renewcommand{\thefootnote}{\arabic{footnote}} \setcounter{footnote}{0}
\noindent

A major observational program is underway to determine the physical 
properties of the universe. A key element of this program is to fully 
characterise the primordial
fluctuations that seeded structure. The availability of precision 
cosmic microwave background (CMB) datasets from satellite 
experiments \cite{wmap_hinshaw_kogut} and smaller scale ground-based 
experiments \cite{small_scale_cmb_data}, complemented by
large-scale structure (LSS) data from galaxy redshift surveys
\cite{2dF_percival,sdss_tegmark}, has now made it possible to work 
towards this ambitious goal.

The simplest characterisation of the primordial perturbations 
involves a Gaussian, nearly scale-invariant spectrum of fluctuations 
arising from adiabatic initial conditions. These features are 
motivated by the simplest single-field models of
inflation. To rigorously establish the character of the primordial 
perturbations, however, it is necessary to study a wider class of 
models. It therefore seems reasonable to consider more general 
possibilities for the primordial perturbations.

In this paper we adopt a phenomenological approach to the problem 
of determining the initial conditions of structure formation. 
Adiabatic initial conditions, characterised by a single, spatially 
uniform equation of state for the stress-energy content of the 
universe, have so far received much attention in the literature.
Such fluctuations arise from perturbing all particle species in 
spatially uniform ratios, resulting in an overall curvature 
perturbation to the hypersurfaces of constant cosmic temperature. 
However, isocurvature perturbations also provide regular
initial conditions for the evolution of fluctuations. Perturbations 
of this type result from spatially varying abundances of particle 
species that are arranged to cancel locally so that the curvature of 
the spatial hypersurface is unperturbed.

In a universe filled with photons, neutrinos, baryons and a cold 
dark matter (CDM) component, four regular isocurvature modes 
arise in addition to the familiar adiabatic mode. The cold dark 
matter isocurvature (CI) and the baryon
isocurvature (BI) modes, in which spatial variations in the cold 
dark matter density or baryon density are compensated by photon 
density perturbations, were studied some time ago 
\cite{bond_efsth_cdmi_and_peebles_bi}. The possibility of
creating a neutrino isocurvature density (NID) mode, in which 
perturbations in the neutrino density are balanced by opposing 
photon density fluctuations, or a neutrino isocurvature velocity 
(NIV) mode, where perturbations in the velocity of
neutrinos are compensated by equal and opposite photon velocity 
perturbations, was realised more recently 
\cite{bmt_2000,rebhan_schwarz_and_lasenby_challinor}.

When two or more perturbation modes are excited, the possibility 
of non-trivial correlations between these modes arises.
The most general Gaussian perturbation of the five
regular modes is completely characterised by the symmetric, 
matrix-valued power spectrum, 
$$P_{ij}(k)\cdot\delta^3({\bf k} - {\bf k}') = \langle A_i({\bf
k})\,A_j({\bf k}')\rangle\, ,$$ 
where the indices $(i,j=1,2,3,4,5)$ label the
modes and $A_i$ are the mode amplitudes \cite{bmt_2000}. If 
attention is restricted to quadratic observables, like the 
CMB angular power spectrum, this description also suffices for 
non-Gaussian fluctuations. Under the assumption that the power 
spectrum for each mode is a smoothly varying function of $k$, each
auto-correlation and cross-correlation mode may be parameterised 
by an amplitude and a spectral index.

The viability of isocurvature perturbations in
the light of observational data has been explored before
\cite{lit_iso,trotta_etal}, over restricted sets of parameters.
A detailed exploration of the likelihood space of cosmological 
parameters and initial
conditions was undertaken in \cite{bmt_2002} to forecast how 
the WMAP and
Planck satellite experiments would measure the amplitude of 
primordial isocurvature fluctuations. However, due to the 
unavailability of satellite data, this analysis was restricted to 
a perturbative exploration of the likelihood surface around an 
adiabatic model. 
Here we utilize current CMB anisotropy and large-scale structure data 
to measure the amplitude of correlated non-adiabatic fluctuations 
over a broad range of flat cosmologies.

\begin{figure}[t!]
\epsfig{file=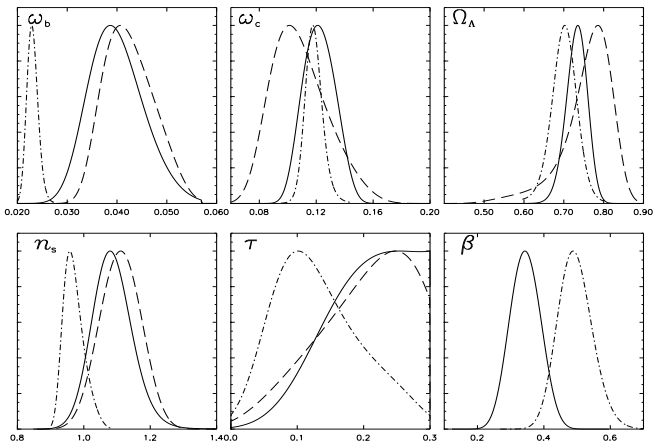,width=8.5cm,height=6cm}
\vskip +0.05in
\caption{Marginalized distributions for the `cosmological' quantities,
for CMB (dashed), CMB+LSS (solid). Results for pure adiabatic
models with CMB+LSS are dot-dashed.}
\label{cosmo_hist}
\vskip -0.2in
\end{figure}

We consider a sixteen dimensional parameter space that includes 
the physical baryon
and cold dark matter density parameters, $\omega_b$ and $\omega_c,$ 
the fractional energy density of a cosmological constant, 
$\Omega_\Lambda,$ the
optical depth parameter to the reionisation epoch, $\tau,$ and $\beta,$
quantifying redshift-space distortions in the
large-scale structure measurements.  
As a prior, we require $\tau \leq 0.3,$ as in
\cite{wmap_spergel}.  The mode correlations are described by the ten 
parameters, $\langle A_i\,A_j\rangle,$ and a single overall tilt 
parameter, $n_s,$ for all the modes. 
To avoid introducing an artificial degeneracy, 
we exclude the BI mode, as its prediction for the CMB
spectrum is identical to that of the CI mode \cite{bmt_2002}.

We treat the mode contributions as follows. Firstly, the CMB and 
matter power spectra for each pure mode are normalized by its 
mean square power, defined as its contribution,\, 
$\langle \delta T^2 \rangle = \sum(2\ell+1)\,C_\ell,\,$ to
the CMB temperature anisotropy from $\ell_{min}=2$ to 
$\ell_{max}=2000.$ The cross-correlation spectra are rescaled 
by the geometric mean of the corresponding pure mode powers. A 
mixed model is constructed from these spectra using
coefficients, $z_{ij},$ that are normalized to have total {\it rms} 
amplitude equal to one. Geometrically a set of symmetric 
$z_{ij}$ corresponds to a position ${\bf z},$ on the
unit nine-sphere. The CMB and matter power spectra for the 
mixed model are then rescaled by the mean square power of the 
admixture before being multiplied by a normalization parameter, 
$P$, to yield the final model spectra. The parameter $P$ quantifies 
the mean square power of the resulting model, which is well 
measured from current data. We impose uniform priors on the 
direction ${\bf z},$ and on $P$.

The CMB temperature and temperature-polarisation cross-correlation 
angular power spectra, and the matter power spectrum were 
evaluated using a modified version of the
grid-based DASh software package \cite{dash}, extended to compute
auto-correlation and cross-correlation spectra for adiabatic 
and isocurvature modes. The computation of 10 mode spectra, for 
a given cosmological model, takes 20 seconds on a single Pentium 
3 processor. The CMB power spectra computed with our improved 
version of DASh agree with those computed using the latest version of
CMBFAST \cite{cmbfast} to within 0.5\%, which is sufficient for 
our likelihood computations.

To sample the parameter space efficiently, we employ a 
Markov Chain Monte Carlo method, using the Metropolis algorithm. 
We are able to sample the posterior distribution for each parameter 
quickly and accurately using an efficient proposal
distribution, for which we take a scaled best guess of the 
covariance matrix of the posterior. To obtain an adequate 
estimate of this covariance matrix,
requires updating a series of chains, of cumulative length 
$\approx 10^5$ steps. However, with this approximate covariance matrix, 
the subsequent single chains used in our analysis converge over 
much shorter lengths, requiring typically 20,000 likelihood 
calculations. Our criterion for
convergence to the underlying distribution is that the sample 
variance of the mean of each parameter, computed using a 
spectral method, is less than one percent of the variance of 
that parameter. More details of the above methods are given in
\cite{bdfms_2}.

\begin{figure}[t!]
\epsfig{file=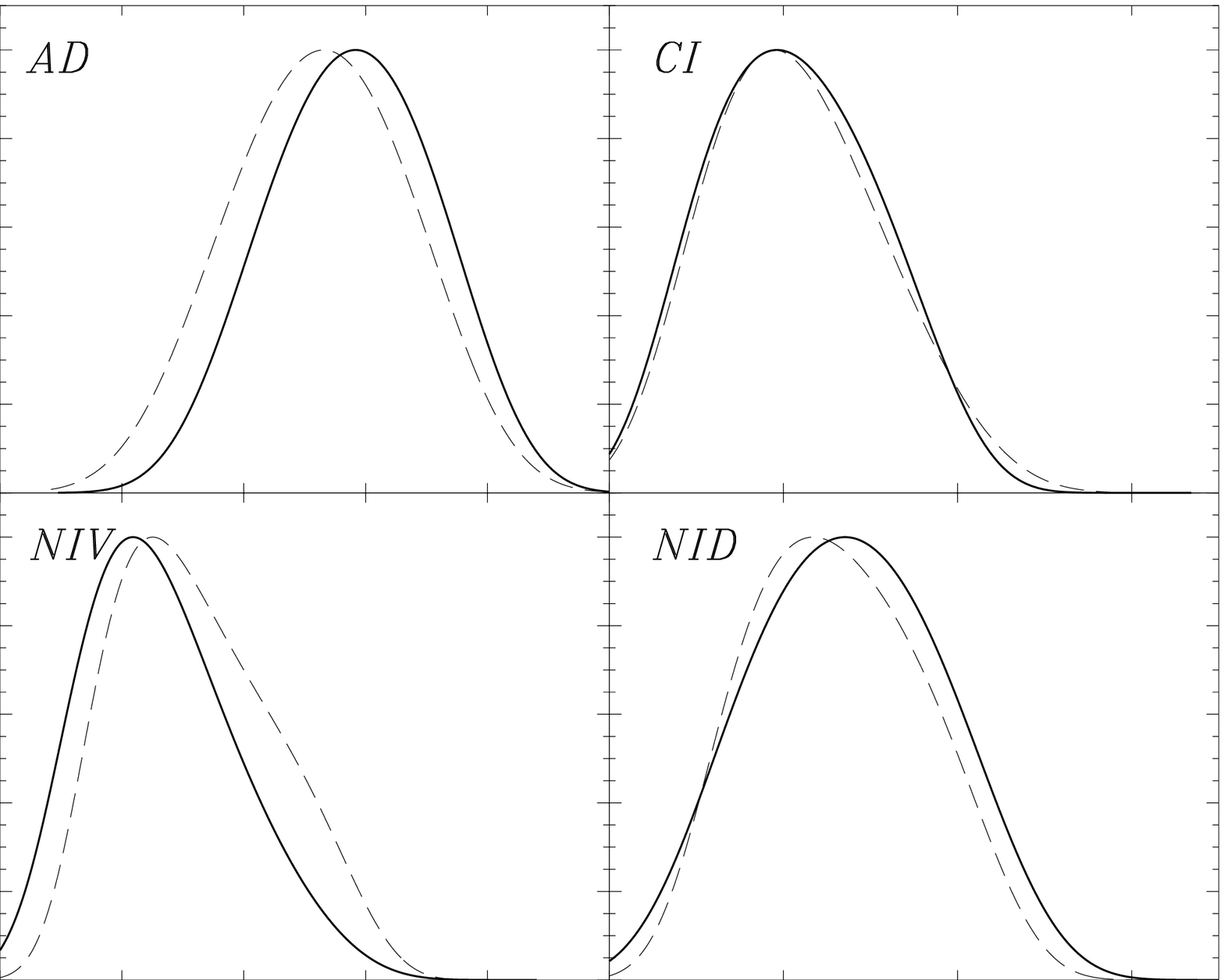,width=8.5cm,height=5cm}
\vskip +0.2in
\epsfig{file=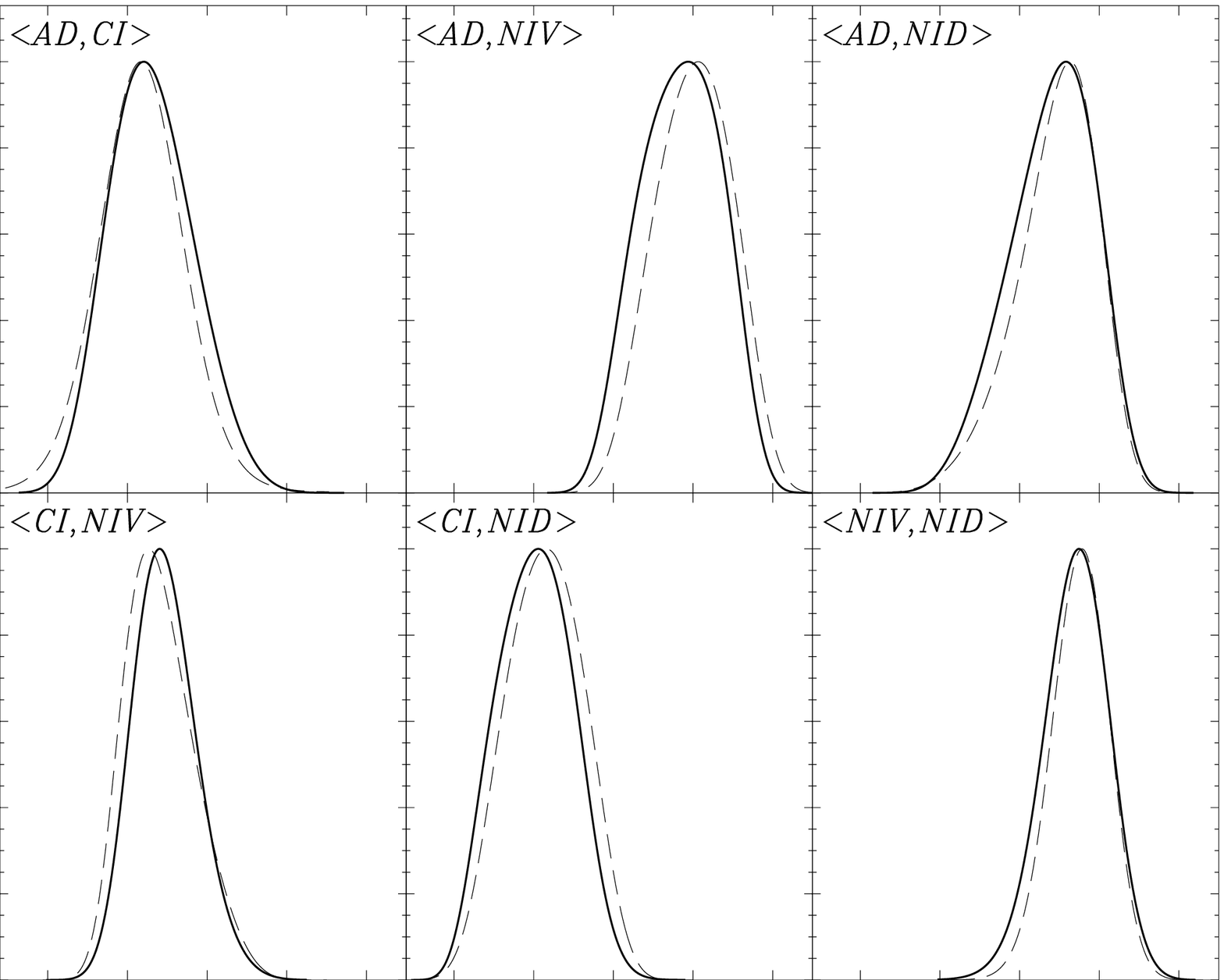,width=8.5cm,height=5cm}
\vskip +0.2in
\label{amp_auto_hist}
\caption{Top four panels: marginalized distributions of relative 
power contributions $z_{ii}$ of the four auto-correlated modes: 
adiabatic (AD), cold dark matter
density isocurvature (${\rm CI}$), neutrino velocity
isocurvature (${\rm NIV}$) and neutrino density isocurvature
(${\rm NID}$) for CMB (dashed) and CMB+LSS (solid).
Bottom six panels: marginalized distributions for relative power 
contributions $z_{ij}$ of cross-correlated modes, for CMB (dashed) 
and CMB+LSS (solid).} 
\vskip -0.2in
\end{figure}

We compare our model spectra to measurements of the CMB anisotropy 
from WMAP \cite{wmap_hinshaw_kogut} and a compilation of CMB data 
on small scales \cite{tegmark_compilation} from the ACBAR, BOOMERANG, 
CBI, and VSA experiments \cite{small_scale_cmb_data}. A combination 
of the galaxy power spectrum measurement by 2dFGRS \cite{2dF_percival} 
and CMB measurements constitutes our
second dataset.  We use the CMB likelihood functions provided 
in \cite{verde} and \cite{tegmark_compilation} for the WMAP and 
small-scale CMB data, respectively, and
the large-scale structure likelihood function available in 
\cite{2dF_percival}, restricting ourselves to the linear regime 
$0.01 < k < 0.15~h {\text{Mpc}}^{-1}$.
The normalization of the observed galaxy power spectrum depends 
on the redshift-space distortion parameter, $\beta,$ that was 
measured in \cite{peacock} to be \,$0.43 \pm 0.07$, which we include 
as a Gaussian prior. We follow the method
of \cite{verde} in normalizing the theoretical matter power spectrum 
to the galaxy power spectrum. However, we do not include an 
independent prior on the bias, $b,$ nor treat the effects of peculiar 
velocities, important at smaller scales.

\begin{figure}[t!]
\epsfig{file=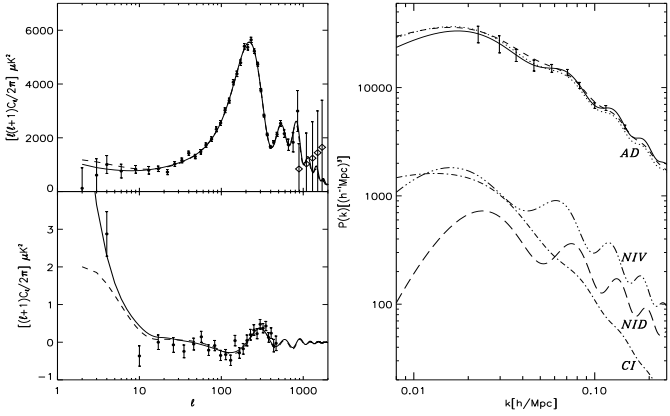,width=9cm,height=5cm}
\vskip 0.25in 
\caption{The CMB temperature spectrum (panel 1) and the
temperature-polarization cross-correlation  
spectrum (panel 2) are plotted for
the maximum likelihood adiabatic model (dashed line) with 
parameter values
($\omega_b$,\,$\omega_c$,\,$\Omega_\Lambda$,\,$n_s$,\,$\tau$,\,$\beta$)$
=$(0.023,\,0.12,\,0.72,\,0.97,\,0.14,\,0.49) and a high-likelihood 
mixed model (solid line) with parameter values 
(0.041,\,0.13,\,0.75,\,1.06,\,0.28,\,0.37), with 
the CMB data overplotted. The galaxy power spectrum $P(k)$ for these
models along with the pure mode contributions to the mixed model is shown
in panel 3, including sizes of LSS error bars.}
\label{cl_plot}
\vskip -0.2in
\end{figure}

For pure adiabatic models our results are consistent with the WMAP 
analysis \cite{wmap_spergel}, for CMB data alone and for CMB data 
combined with LSS data.  
The quality of fit of the highest likelihood models, together 
with the median parameter values and 68\% confidence intervals are 
given in table \ref{table1}.  
The marginalised parameter distributions are also displayed in
Fig.~\ref{cosmo_hist}. When correlated adiabatic and isocurvature initial
conditions are allowed the situation changes significantly. We find 
that large fractions of isocurvature models are tolerated by current 
data. The best-fit mixed model has a likelihood higher by a 
factor $\approx e^{5/2}$ than that of the best-fit pure adiabatic model.
Fig.~\ref{cl_plot} shows the best-fit model spectra with their parameter
values listed in the caption; the pure adiabatic model and the mixed 
model are
essentially indistinguishable. The relative magnitude of the 
non-adiabatic component $z_{\mbox{\tiny ISO}}$, defined
as $z_{\mbox{\tiny ISO}}=\sqrt {(1-z_{\mbox{\tiny AD}}^2)},$ is 
measured for the CMB data alone to be 
$0.84^{+0.08}_{-0.13}\,$  and
comprises roughly equal proportions of pure isocurvature modes. To 
extract a fractional non-adiabatic contribution  we define 
$f_{\mbox{\tiny ISO}}= 
z_{\mbox{\tiny ISO}}/(z_{\mbox{\tiny ISO}}+z_{\mbox{\tiny AD}})$, 
calculated to be $0.60^{+0.09}_{-0.11}\,$.

If large-scale structure data is included the relative magnitude of
non-adiabatic modes is slightly reduced to the median value of $0.79
^{+0.09}_{-0.13}\,$, with $f_{\mbox{\tiny ISO}}=0.57\pm0.09$. The 
inclusion of isocurvature modes does not improve the fit
to the 2dF data, but neither does this dataset significantly 
constrain the isocurvature fraction. This is true because we have 
allowed the normalization of the galaxy power spectrum to vary, only 
indirectly constraining it through our
prior on $\beta$. We find that smaller values of $\beta$ are favoured,
corresponding to larger biases at fixed $\Omega_m$. This is not 
surprising given that the matter power spectra generated in 
isocurvature models are lower in amplitude relative to the adiabatic 
spectrum, as illustrated in Fig. \ref{cl_plot}. Therefore
models with higher isocurvature content have correspondingly larger 
biases. For both datasets the median value of $n_s$ is larger than in 
the adiabatic case with a much broader distribution.

Turning to the cosmological parameters, we find that $\omega_c$ and
$\Omega_\Lambda$ have broader distributions shifted relative to the
adiabatic case, when CMB data alone is considered.  The constraints 
on these parameters become tighter when 2dF data is included, mainly 
because the matter power spectrum is sensitive to a combination of 
$\Omega_m$ and $h$, the dimensionless Hubble parameter, which in our 
parameterisation translates into improved constraints on $\omega_c$ 
and $\Omega_\Lambda$, within the range of flat
models. The reionisation parameter, $\tau,$ which has a fairly broad 
distribution in the adiabatic case, is very poorly constrained in 
mixed models.

\begin{table}[t!]
\vspace{-.2in}
\begin{center}
\begin{tabular}{lcccc}
\hline\hline
& \hspace{-8mm}\mbox{\fns ADIA}&  \hspace{-4mm} \mbox{\fns ADIA} & 
\mbox{\fns ADIA+ISO} & \mbox{\fns ADIA+ISO}\\
& \hspace{-8mm}\mbox{\fns CMB} & \hspace{-4mm} \mbox{\fns CMB+LSS} 
& \mbox{\fns CMB} & \mbox{\fns CMB+LSS} \\ \hline
%%--------------------------------------------------------------%%
\hspace{0.25cm}$\omega_b$ & \hspace{-10mm} \mbox{\fns $0.024\pm 0.001$} & 
\hspace{-4mm} \mbox{\fns $0.023\pm 0.001$} &
\mbox{\fns $ 0.043\pm 0.005$}  & \mbox{\fns $0.041\pm 0.006$} \\
%%--------------------------%%-----------------%%----------------%%
\hspace{0.25cm}$\omega_c$ & \hspace{-10mm}\mbs{$0.13\pm 0.01$}
& \hspace{-4mm} \mbox{\fns $0.120\pm 0.006$} & \mbs{$0.11\pm0.02$} 
& \mbs{$0.12\pm0.01$} \\  
%%--------------------------%%-----------------%%----------------%%
\hspace{0.25cm}$\Omega_\Lambda$ 
& \hspace{-10mm}\mbs{$0.69\,$}$^{+\,0.06}_{-\,0.08}$
& \hspace{-4mm} \mbs{$0.71\pm 0.03$} & 
\mbs{$0.79\,$}$^{+\,0.05}_{-\,0.07}$  & \mbs{$0.74\pm0.03$} \\
%%--------------------------%%-----------------%%----------------%%
 \hspace{0.25cm}$n_s$ & \hspace{-10mm}\mbs{$0.99\pm0.03$}
& \hspace{-4mm} \mbs{$0.97\pm0.03$} & 
\mbs{$1.13\pm0.07$}  & \mbs{$1.10\pm0.06$} \\
%%--------------------------%%-----------------%%----------------%%
 \hspace{0.25cm}$\tau$ & \hspace{-10mm}\mbs{$0.15\pm0.07$} & \hspace{-4mm}
\mbs{$0.13\,$}$^{+\,0.08}_{-\,0.06}$ & 
\mbs{$0.21\,$}$^{+\,0.06}_{-\,0.08}$ &  \mbs{$0.22\pm0.07$} \\
%%--------------------------%%-----------------%%----------------%%
 \hspace{0.25cm}$\beta$ & \hspace{-11mm}$ \hld $
& \hspace{-3mm}\mbs{$0.50\pm0.05$} & $\hld $ & \mbs{$0.35\pm0.05$} \\
%%--------------------------%%-----------------%%----------------%%
 $\langle \mbox{\fns AD, AD} \rangle$ & \hspace{-10mm} \mbs{1.0} & 
\hspace{-4mm} \mbs{1.0} & \hspace{-1mm}
\mbs{$0.55\,$}$^{+\,0.16}_{-\,0.14}$ & \mbs{$0.61\pm{0.15}$} \\
%%--------------------------%%-----------------%%----------------%%
$\langle \mbox{\fns CI, CI} \rangle$ & \hspace{-11mm}$\hld$ & 
\hspace{-6mm} $ \hld $ & 
\mbs{$0.23\,$}$^{+\,0.11}_{-\,0.09}$ & \mbs{$0.23\pm0.11$} \\
%%--------------------------%%-----------------%%----------------%%
$\langle  \mbox{\fns NIV, NIV}\rangle$ & \hspace{-11mm}$\hld$ & 
\hspace{-5mm}$ \hld $ & 
\mbs{$0.34\,$}$^{+\,0.14}_{-\,0.13}$ & \mbs{$0.28\,$}$^{+\,0.14}_{-\,0.11}$ \\
%%--------------------------%%-----------------%%----------------%%
$\langle \mbox{\fns NID, NID} \rangle$ & \hspace{-11mm}$\hld$ & 
\hspace{-6mm} $ \hld $ & 
\mbs{$0.28\,$}$^{+\,0.12}_{-\,0.10}$ & \mbs{$0.30\pm0.12$} \\
%%--------------------------%%-----------------%%----------------%%
$\langle \mbox{\fns AD, CI} \rangle$ & \hspace{-11mm}$\hld$ & 
\hspace{-6mm} $ \hld $ &
 \hspace{-4mm}
 \mbs{$-0.14\pm0.10$} &\hspace{-3mm} 
 \mbs{$-0.12\,$}$^{+\,0.12}_{-\,0.10}$ \\
%%--------------------------%%-----------------%%----------------%%
$\langle \mbox{\fns AD, NIV} \rangle$ & \hspace{-11mm}$\hld$ & 
\hspace{-6mm} $ \hld $ & 
\mbs{$0.22\pm0.10$} & \mbs{$0.19\pm0.11$} \\
%%--------------------------%%-----------------%%----------------%%
$\langle \mbox{\fns AD, NID} \rangle$ & \hspace{-11mm}$\hld$ & 
\hspace{-6mm} $ \hld $ & 
\mbs{$0.12\,$}$^{+\,0.09}_{-\,0.11}$ & \mbs{$0.11\,$}$^{+\,0.10}_{-\,0.12}$ \\
%%--------------------------%%-----------------%%----------------%%
$\langle \mbox{\fns CI, NIV} \rangle$ & \hspace{-12mm} $\hld$ & 
\hspace{-5mm}$ \hld $ & 
 \hspace{-4mm}
\mbs{$-0.10\,$}$^{+\,0.10}_{-\,0.08}$ & \hspace{-2.5mm}\mbs{$-0.09\pm0.08$} \\
%%--------------------------%%-----------------%%----------------%%
${\langle \mbox{\fns CI, NID} \rangle}$ & \hspace{-12mm} $\hld$ &
\hspace{-6mm} $ \hld $ &
 \hspace{-4mm}
\mbs{$-0.15\pm0.10$} &  \hspace{-3.5mm} \mbs{$-0.18\pm0.10$} \\
%\mbs{$-0.15\pm0.10$} & \mbs{$-0.18\,$}$^{+\,0.09}_{-\,0.10}$ \\
%%--------------------------%%-----------------%%----------------%%
$\langle \mbox{\fns NIV, NID} \rangle$ & \hspace{-11mm}$\hld$ & 
\hspace{-6mm} $ \hld $ & 
\mbs{$0.17\pm0.07$} & \mbs{$0.16\pm0.08$} \\  
%%--------------------------%%-----------------%%----------------%%
\hline 
\hspace{0.25cm} $\chi^2/\nu$ &  \mbs{$1435/1347$} & \mbs{$1463/1378$} & 
\mbs{$1430/1338$} & \mbs{$1458/1369$} \\
\hline
\end{tabular}
\end{center}
\caption{Median parameter values, $68\%$ confidence intervals, 
and best fit $\chi^2$ per degree of freedom $\nu$, for pure adiabatic 
models (columns 1 and 2) and mixed models (columns 3 and 4). The coefficients 
$z_{ij}$ are tabulated for the different modes.}

\label{table1}
\vskip -0.2in
\end{table}

\begin{figure}[t!]
\vskip -0.1in
\epsfig{file=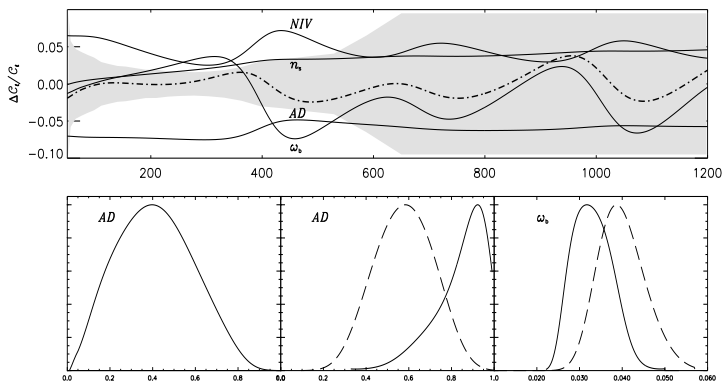,width=8.5cm,height=5.3cm}
\vskip 0.1in
\caption{Top panel: The total derivative along the degenerate 
direction (dot-dashed) described in the text is decomposed into 
variations in four parameters ($\omega_b$, $n_s$, AD, NIV), which 
cancel to within the experimental error bars (shaded). Bottom panel: 
(left) marginalized prior for 
the adiabatic mode contribution $z_{\mbox{\tiny AD}}$, no data; (centre) 
marginalized posterior for $z_{\mbox{\tiny AD}}$ with uniform prior 
on all models (dashed) and with uniform prior on $z_{\mbox{\tiny AD}}$ 
(solid); (right) 
marginalized posterior for $\omega_b$ with priors as in centre panel.}
\label{degen}
\vskip -0.2in
\end{figure}

The greatest impact, however, is on the baryon density distribution, 
which is significantly broadened, with a median value twice as large 
as in the adiabatic case. The drastic shift in the baryon density is 
associated with a flat direction in parameter space, which was 
observed previously in \cite{trotta_etal} but not
investigated further. Here we clarify the nature of this degeneracy. In
Fig.~\ref{degen} we illustrate how a change in a high-likelihood model 
$C_\ell$ due to a 10\% increase in $\omega_b$ is compensated by a 
40\% increase in the power of the pure NIV mode, a 1.4\% increase in 
the value of $n_s$ and a 14\% decrease in the power of the pure AD mode, 
to well within the CMB error bars. We have found that there is a 
high-likelihood plateau connecting the mixed
model to the adiabatic model. It is along this plateau that 
$\omega_b$ and $n_s$ are shifted to higher values.

We found that the allowed fractional contribution of isocurvature
modes depends sensitively on the choice of prior. The problem is
that no natural notion of a ``uniform" prior stands out, in 
particular when correlations between the modes are allowed. The prior 
presented earlier can be argued to be biased
against models where a single mode dominates (such as the pure adiabatic
model) because 
such models 
have a comparatively minuscule phase space density. This 
``entropy" effect is illustrated in the bottom left hand panel of 
Fig. \ref{degen}, where the {\it a posteriori} 
distribution for $z_{\mbox{\tiny AD}}$ that would result with no data
is plotted. 
We observe that the pure adiabatic model has zero 
density solely on account of phase space volume effects. In the bottom 
centre panel of Fig. \ref{degen}
we show the posterior distribution for $z_{\mbox{\tiny AD}}$ that would 
result from dividing out this phase space density effect (solid curve),  
compared to the original posterior (dashed). 
After the rescaling of the prior, 
the allowed isocurvature contribution is greatly suppressed. In 
the bottom right panel of Fig. \ref{degen} we see that 
the reweighted posterior for $\omega_{b}$ overlaps with the conventional 
value, although much higher values are still allowed.

We have restricted our analysis to flat models and have assigned 
a single spectral index to all modes. Relaxing either of these 
restrictions will open up interesting regions of parameter space. 
However, it is evident from the results
presented here that further precision data is required before meaningful 
constraints can be placed on extended parameter sets. Future 
high quality CMB polarisation data promises to provide such strong 
constraints \cite{bmt_2001}, 
while improved temperature and polarisation spectra from WMAP, the 
recently available SDSS dataset and higher quality small-scale 
temperature measurements from ground-based experiments will allow 
further progress on these issues. 

{\it Acknowledgments}:  MB thanks 
Mr D.~Avery for support through the SW Hawking Fellowship in
Mathematical Sciences at Trinity Hall. JD acknowledges a PPARC 
studentship. PGF thanks the Royal Society. KM acknowledges a 
PPARC Fellowship and a Natal University research grant. CS is 
supported by a Leverhulme Foundation grant. 
We thank  R.~Durrer, U.~Seljak, D.~Spergel and R.~Trotta for 
useful discussions.

\tighten

\end{document}